\documentclass[twocolumn,showpacs,preprintnumbers,amsmath,amssymb]{revtex4}
\usepackage{graphicx}
\usepackage{dcolumn}
\usepackage{bm}
\begin{document}
\title{Testing the littlest Higgs model with $T$ parity in bottom quark pair
production at high energy photon colliders}
\author{Jinshu Huang}
\email{jshuang@vip.sina.com} \affiliation{College of Physics $\&$
Information Engineering, Henan Normal University, Xinxiang 453007,
P. R. China;
\\College of Physics $\&$ Electric Engineering, Nanyang Normal
University, Nanyang 473061, P. R. China; \\ Kavli Institute for
Theoretical Physics China, Chinese Academy of Science, Beijing
100190, P. R. China}
\author{Gongru Lu}
\email{lugongru@sina.com}
\author{Xuelei Wang}
\email{wangxuelei@sina.com} \affiliation{College of Physics $\&$
Information Engineering, Henan Normal University, Xinxiang 453007,
P. R. China; \\ Kavli Institute for Theoretical Physics China,
Chinese Academy of Science, Beijing 100190, P. R. China}

\date{\today}

\begin{abstract}
We have calculated the cross section of the process $e^+e^-
\rightarrow \gamma\gamma \rightarrow b\bar{b}$ in the littlest Higgs
model with $T$ parity (LHT). We find that, for the favorable
parameters, the total cross section $\sigma(e^+e^- \rightarrow
\gamma\gamma \rightarrow b\bar{b})$ is sensitive to the breaking
scale $f$, mixing parameter $x_L$, the masses of the mirror quarks
$m_{Hi}$, and the relative correction of the LHT model is a few
percent to dozens of percent. The cross section is significantly
larger than the corresponding results in the standard model and in
the other typical new physics models. Therefore the prediction in
the LHT model is quite different from the predictions in other new
physics models and such a process is really interesting in searching
for the signs of the LHT model.
\end{abstract}


\maketitle

\section{\label{sec:level1}Introduction}

The electroweak symmetry breaking mechanism remains an open question
in spite of the success of the standard model (SM) compared with the
precision measurement data. The collisions of high energy photons
produced at the linear collider provide a comprehensive laboratory
for testing the SM and probing new physics beyond the SM
\cite{Brodsky1995}. With the advent of the new collider technique,
the high energy and high intensity photon beams can be obtained by
using Compton laser photons scattering off the colliding electron
and positron beams \cite{Ginzburg1981}, and a large number of heavy
quark pairs can be produced by this method. The photon energy
spectrums show that there are many relatively soft photons, and the
production of heavy top quark will be suppressed owing to the
reduction of collision energies. However, no such suppression
affects the relatively light bottom quark \cite{Halzen1992}.
Therefore it is worth investigating the production of the bottom
quark pairs in the photon-photon collisions.

In the SM, this process has been calculated and the QCD threshold
effects of the process have been also examined \cite{Eboli1993}.
Reference \cite{Han1996} presents a study of the Yukawa corrections
to this process in both the general two Higgs doublet model (2HDM)
and the minimal supersymmetric standard model (MSSM), which arise
from the virtual effects of the charged Higgs and charged Goldstone
bosons, and shown that the relative correction to the total cross
section of the processes $e^+e^- \rightarrow \gamma\gamma
\rightarrow b\bar{b}$ is less than $0.1\%$ for the favorable
parameter values. In Ref. \cite{Huang2008}, the authors have
calculated the Yukawa correction to the cross section of
$\gamma\gamma \rightarrow b\bar{b}$ induced by the pseudo-Goldstone
bosons and the new gauge bosons in the topcolor assisted technicolor
(TC2) model, and pointed out that the relative correction is
negative and not more than $10\%$. In this paper, we will study the
contribution of the littlest Higgs model with $T$ parity (LHT) to
this process.

As we know, the fancy idea of little Higgs \cite{Georgi2001} tries
to provide an elegant solution to the hierarchy problem by regarding
the Higgs boson as a pseudo-Goldstone boson, whose mass is protected
by an approximate global symmetry, and the quadratic divergence
cancellation is due to the contributions from new particles with the
same spin as the SM particles. The littlest Higgs model
\cite{Arkani2002} is a cute economical implementation of the little
Higgs idea, but is found to be subject to the strong constraints
from electroweak precision tests \cite{Csaki2003}, which would
require raising the mass scale of the new particles to far above TeV
scale and thus reintroduce the fine-tuning in the Higgs potential
\cite{Marandella2005}. To tackle this problem, a discrete symmetry
called $T$ parity is proposed \cite{Cheng2003}, which forbids the
tree-level contributions from the heavy gauge bosons to the
observables involving only the SM particles as external states.
Therefore we will investigate the process $\gamma\gamma \rightarrow
b\bar{b}$ in this model.

This paper is organized as follows. In Sec. II, we present a brief
review of the LHT model. Section III is devoted to our analytical
results of the cross section of $e^+e^- \rightarrow \gamma\gamma
\rightarrow b\bar{b}$ in terms of the well-known standard notation
of one-loop Feynman integrals. The numerical results and conclusions
are included in Sec. IV.

\section{\label{sec:level2} A brief review of the LHT model}

The LHT model \cite{Cheng2003,Hubisz2006,Blanke2006} is based on a
nonlinear sigma model describing the spontaneous breaking of a
global $SU(5)$ down to a global $SO(5)$ at the scale $f \sim O(\rm
TeV)$. From the $SU(5)/SO(5)$ breaking, there arise $14$
Nambu-Goldstone bosons which are described by the ``pion'' matrix
$\Pi$, given explicitly by
\begin{widetext}
\begin{equation}
\Pi=\left ( \begin{array}{ccccc}
-\frac{\omega^0}{2}-\frac{\eta}{\sqrt{20}} &
-\frac{\omega^+}{\sqrt{2}} & -i\frac{\pi^+}{\sqrt{2}} & -i\phi^{++}
& -i\frac{\phi^+}{\sqrt{2}}\\
-\frac{\omega^-}{\sqrt{2}} &
\frac{\omega^0}{2}-\frac{\eta}{\sqrt{20}} & \frac{v+h+i\pi^0}{2} &
-i\frac{\phi^+}{\sqrt{2}} & \frac{-i\phi^0+\phi^P}{\sqrt{2}} \\
i\frac{\pi^-}{\sqrt{2}} & \frac{v+h-i\pi^0}{2} & \sqrt{4/5}\eta &
-i\frac{\pi^+}{2} & \frac{v+h+i\pi^0}{2} \\
i\phi^{--} & i\frac{\phi^-}{\sqrt{2}} & i\frac{\pi^-}{\sqrt{2}} &
-\frac{\omega^0}{2}-\frac{\eta}{\sqrt{20}} &
-\frac{\omega^-}{\sqrt{2}} \\
i\frac{\phi^-}{\sqrt{2}} & \frac{i\phi^0+\phi^P}{\sqrt{2}} &
 \frac{v+h-i\pi^0}{2} & -\frac{\omega^+}{\sqrt{2}} &
 \frac{\omega^0}{2}-\frac{\eta}{\sqrt{20}}
\end{array} \right ).
\end{equation}
\end{widetext}

Under T-parity, the SM Higgs doublet
\begin{equation}
H=\left ( \begin{array}{c}
-i\frac{\pi^+}{\sqrt{2}}\\
\frac{v+h+i\pi^0}{2}
\end{array} \right ),
\end{equation}
is T-even, while the other fields including a physical scalar
triplet
\begin{equation}
\Phi=\left ( \begin{array}{cc}
 -i\phi^{++}
& -i\frac{\phi^+}{\sqrt{2}}\\
-i\frac{\phi^+}{\sqrt{2}} & \frac{-i\phi^0+\phi^P}{\sqrt{2}}
\end{array} \right ),
\end{equation}
and heavy Goldstone bosons $\omega^{\pm}$, $\omega^0$, $\eta$ are
T-odd.

A subgroup $[SU(2)\times U(1)]_1 \times [SU(2)\times U(1)]_2$ of the
$SU(5)$ is gauged, and it is broken into the SM electroweak symmetry
$SU(2)_L \times U(1)_Y$ at the scale $f$. The Goldstone bosons
$\omega^0, \omega^{\pm}$ and $\eta$ are, respectively, eaten by the
new T-odd gauge bosons $Z_H$, $W_H$ and $A_H$, which obtain masses
at the order of $O(v^2/f^2)$
\begin{equation}
M_{W_H} = M_{Z_H} = fg(1-\frac{v^2}{8f^2}), \ \  M_{A_H} =
\frac{fg'}{\sqrt{5}} (1-\frac{5v^2}{8f^2}),
\end{equation}
with $g$ and $g'$ being the SM $SU(2)$ and $U(1)$ gauge couplings,
respectively.

The masses of the SM T-even $Z$ boson and $W$ boson are generated
through eating the Goldstone bosons $\pi^0$ and $\pi^{\pm}$. They
are given by
\begin{equation}
M_{W_L} = \frac{gv}{2}(1-\frac{v^2}{12f^2}),\ \ M_{Z_L}
=\frac{gv}{2\cos\theta_W}(1-\frac{v^2}{12f^2}).
\end{equation}
The photon $A_L$ is also T-even and massless.

In order to cancel the quadratic divergence of the Higgs mass coming
from top loops, an additional T-even quark $T_+$, as a heavy partner
of top quark, is introduced. The implementation of $T$ parity then
requires also a T-odd partner $T_-$. To leading order, their masses
are given by
\begin{eqnarray}
m_{T_+}&=&\frac{f}{v}\frac{m_t}{\sqrt{x_L(1-x_L)}}[1+\frac{v^2}{f^2}(\frac{1}{3}-x_L(1-x_L))],\
\nonumber \\
m_{T_{-}}&=&\frac{f}{v}\frac{m_t}{\sqrt{x_L}}[1+\frac{v^2}{f^2}(\frac{1}{3}-\frac{1}{2}x_L(1-x_L))],
\end{eqnarray}
where $x_L=\lambda^2_1/(\lambda^2_1+\lambda^2_2)$ is the mixing
parameter between the SM top quark and its heavy partner $T_+$
quark, in which $\lambda_1$ and $\lambda_2$ are the Yuwaka coupling
constants in the Lagrangian of the top quark sector. Furthermore,
for each SM quark (lepton), a copy of mirror quark (lepton) with
T-odd quantum number is added in order to preserve the $T$ parity.
We denote them by $u^i_H$, $d^i_H$, $\nu^i_H$, $l^i_H$, where
$i=1,2,3$ are the generation index. In $O(v^2/f^2)$, the masses of
$u^i_H$ and $d^i_H$ satisfy
\begin{eqnarray}
m^u_{Hi}&=&\sqrt{2}\kappa_i f (1-\frac{v^2}{8f^2})\equiv
m_{Hi}(1-\frac{v^2}{8f^2}),\nonumber \\ m^d_{Hi}&=&\sqrt{2}\kappa_i
f \equiv m_{Hi}.
\end{eqnarray}
where $\kappa_i$ are the diagonalized Yukawa couplings of the mirror
fermions.

The mirror fermions induce a new flavor structure and there are four
CKM-like unitary mixing matrices in the mirror fermion sector:
$V_{Hu}$, $V_{Hd}$, $V_{Hl}$ and $V_{H{\nu}}$. These mirror mixing
matrices are involved in the charged-current flavor-changing
interactions between the SM fermions and the T-odd mirror fermions
which are mediated by the T-odd heavy gauge bosons or the Goldstone
bosons. $V_{Hu}$ and $V_{Hd}$ satisfy the relation
\begin{equation}
V^{\dag}_{Hu} V_{Hd}=V_{\rm CKM}.
\end{equation}
Following the Refs. \cite{Hubisz2006} and \cite{Blanke2006},
$V_{Hd}$ is parameterized  with three angles $\theta^d_{12},
\theta^d_{23}, \theta^d_{13}$ and three phases $\delta^d_{12},
\delta^d_{23}, \delta^d_{13}$, and is obtained with the expression
\begin{widetext}
\begin{equation}
V_{Hd}=\left ( \begin{array}{ccc} c^d_{12}c^d_{13} &
s^d_{12}c^d_{13}e^{-i\delta^d_{12}} & s^d_{13}e^{-i\delta^d_{13}}\\
-s^d_{12}c^d_{23}e^{i\delta^d_{12}}-c^d_{12}s^d_{23}s^d_{13}e^{i(\delta^d_{13}-\delta^d_{23})}
&
c^d_{12}c^d_{23}-s^d_{12}s^d_{23}s^d_{13}e^{i(\delta^d_{13}-\delta^d_{12}-\delta^d_{23})}
& s^d_{23}c^d_{13}e^{-i\delta^d_{23}}\\
s^d_{12}s^d_{23}e^{i(\delta^d_{12}+\delta^d_{23})}-c^d_{12}c^d_{23}s^d_{13}e^{i\delta^d_{13}}
&
-c^d_{12}s^d_{23}e^{i\delta^d_{23}}-s^d_{12}c^d_{23}s^d_{13}e^{i(\delta^d_{13}-\delta^d_{12})}
& c^d_{23}c^d_{13}
\end{array} \right ).
\end{equation}
\end{widetext}

\section{\label{sec:level3} The cross section of bottom pair production in photon-photon collision}

\begin{figure*}

\vspace{-2cm}

\includegraphics{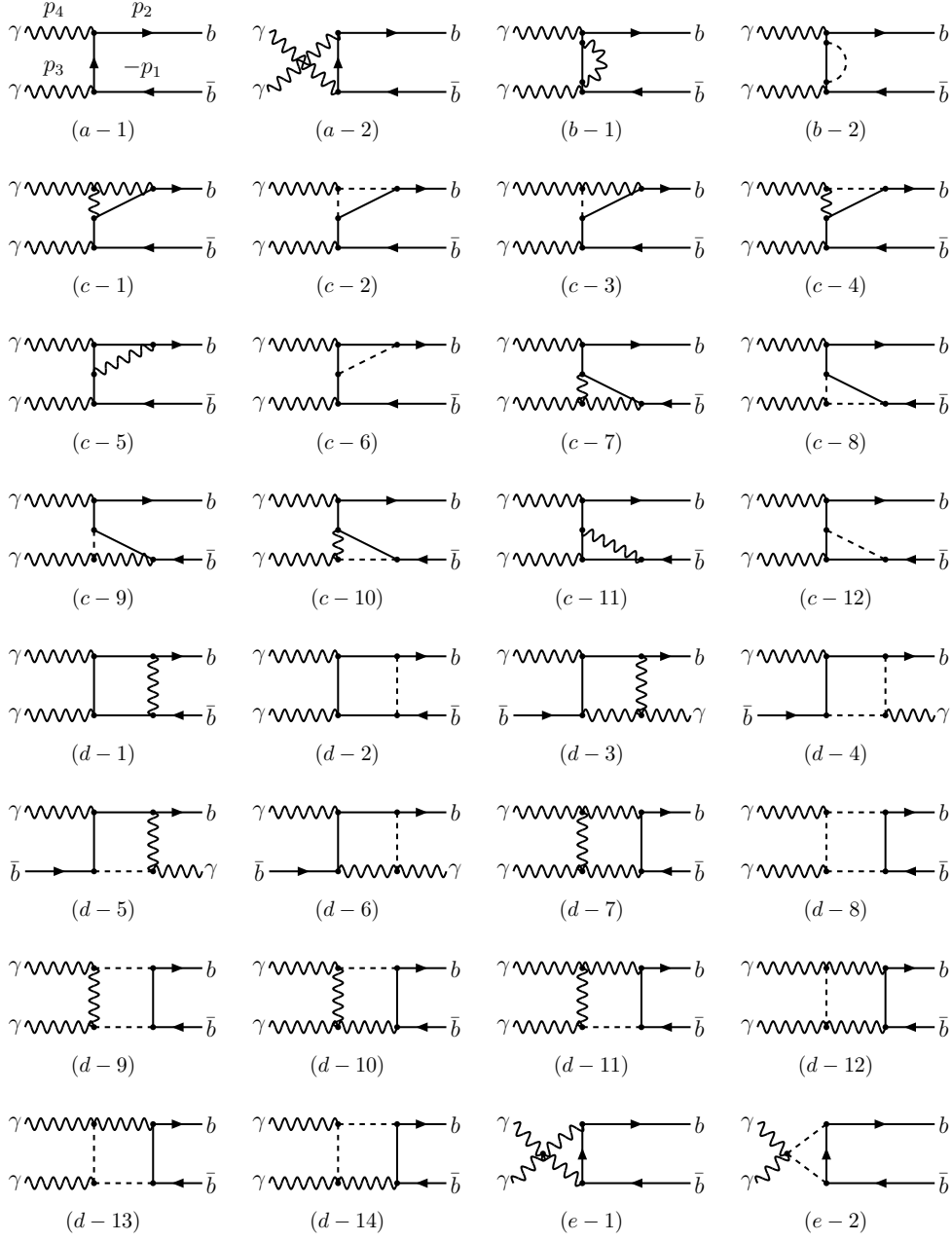}

\vspace{-11cm}

\caption{\label{fig:eps1} Feynman diagrams for the LHT model
contributions to the $\gamma\gamma \rightarrow b\bar{b}$ process:
(a) tree-level diagrams; (b) self-energy diagrams; (c) vertex
diagrams; (d) box diagrams; (e) triangle diagrams. Here only
one-loop diagrams corresponding to the tree-level diagram (a-1)
are plotted. The internal wavy lines represent the gauge bosons
$A_H$, $Z_H$, $W^{\pm}_H$, and $W^{\pm}_L$ in the figures (b-1),
(c-5), (c-11) and (d-1). The dashed lines indicate the Goldstone
bosons $\omega^0$, $\eta$, $\omega^{\pm}$, and $\pi^{\pm}$ in the
figures (b-2), (c-6), (c-12) and (d-2). The internal wavy lines
represent the charged gauge bosons $W^{\pm}_H$ and $W^{\pm}_L$,
together with the dashed lines stand for the charged Goldstone
bosons $\omega^{\pm}$ and $\pi^{\pm}$ in the figures (c-1)-(c-4),
(c-7)-(c-10), (d-3)-(d-14) and (e-1)-(e-2). The internal solid
lines in all the loops denote the fermions $d_H^i$, $u_H^i$, or
$T_+$, which match the corresponding bosons, respectively.}
\end{figure*}

In the LHT model, both T-even and T-odd particles can make the
contributions to the process $\gamma \gamma \rightarrow b\bar{b}$.
The contributions of T-even particles include both the SM
contributions and the contributions of the top quark T-even
partner. The contributions of T-odd particles are induced by the
interactions between the SM quarks and the mirror quarks mediated
by the heavy T-odd gauge bosons or Goldstone bosons. The relevant
Feynman diagrams are shown in Fig.~\ref{fig:eps1}. In our
calculation, we use the dimensional regularization to regulate all
the ultraviolet divergences in the virtual loop corrections, and
adopt the 't Hooft-Feynman gauge and on-mass-shell renormalization
scheme \cite{Bohm1986}. The renormalized amplitude for $\gamma
\gamma \rightarrow b\bar{b}$ contains

\begin{eqnarray}
M_{\rm ren}&=& M_0+ \delta M \nonumber \\ &=& M_0+ \delta M^{\rm
self} + \delta M^{\rm vertex}+ \delta M^{\rm box} \nonumber \\ && +
\delta M^{\rm tr},
\end{eqnarray}
where $M_0$ is the amplitude at the tree level, $\delta M^{\rm
self}$, $\delta M^{\rm vertex}$, $\delta M^{\rm box}$ and $\delta
M^{\rm tr}$ represent the contributions arising from the
self-energy, vertex, box, and triangle diagrams, respectively. Their
explicit forms are given by
\begin{eqnarray}
M_0=M_0^{\hat{t}}+M_0^{\hat{u}},
\end{eqnarray}
\begin{eqnarray}
\delta M^{\rm self}=\delta M^{s(\hat{t})}+\delta M^{s(\hat{u})},
\end{eqnarray}
\begin{eqnarray}
\delta M^{\rm vertex}=\delta M^{v(\hat{t})}+\delta M^{v(\hat{u})},
\end{eqnarray}
\begin{eqnarray}
\delta M^{\rm box}=\delta M^{b(\hat{t})}+\delta M^{b(\hat{u})},
\end{eqnarray}
where
\begin{eqnarray}
M_0^{\hat{t}}&=&-i\frac{e^2 Q_b^2}{\hat{t}-m_b^2} \epsilon_{\mu}
(p_4) \epsilon_{\nu}(p_3) \bar{u}(p_2) \nonumber \\ && \times
\gamma^{\mu} (\not p_2 -\not p_4+m_b) \gamma^{\nu} v(p_1),
\end{eqnarray}

\begin{eqnarray}
M_0^{\hat{u}}=M_0^{\hat{t}}(p_3 \leftrightarrow p_4, \hat{t}
\leftrightarrow \hat{u}),
\end{eqnarray}
\begin{eqnarray}
\delta M^{s(\hat{t})}&=&i\frac{e^2 Q_b^2}{(\hat{t}-m_b^2)^2}
\epsilon_{\mu} (p_4) \epsilon_{\nu}(p_3) \bar{u}(p_2)
[f_1^{s(\hat{t})} \gamma^{\mu} \gamma^{\nu} \nonumber \\ &&+
f_2^{s(\hat{t})} p_2^{\mu} \gamma^{\nu}+f_3^{s(\hat{t})} \not p_4
\gamma^{\mu} \gamma^{\nu}] v(p_1),
\end{eqnarray}
\begin{eqnarray}
\delta M^{s(\hat{u})}=\delta M^{s(\hat{t})} (p_3 \leftrightarrow
p_4, \hat{t} \leftrightarrow \hat{u}),
\end{eqnarray}
\begin{eqnarray}
\delta M^{v(\hat{t})}&=&-i\frac{e^2 Q_b}{\hat{t}-m_b^2}
\epsilon_{\mu} (p_4) \epsilon_{\nu}(p_3) \bar{u}(p_2)
[f_1^{v(\hat{t})} \gamma^{\mu} \gamma^{\nu} \nonumber \\&&+
f_2^{v(\hat{t})} \gamma^{\mu} p_1^{\nu}+ f_3^{v(\hat{t})} p_2^{\mu}
\gamma^{\nu}+
f_4^{v(\hat{t})} p_2^{\mu} p_1^{\nu} \nonumber\\
&&+f_5^{v(\hat{t})} \not p_4 \gamma^{\mu} \gamma^{\nu}
+f_6^{v(\hat{t})} \not p_4 \gamma^{\mu} p_1^{\nu} \nonumber
\\ &&+f_7^{v(\hat{t})} \not p_4 p_2^{\mu} \gamma^{\nu}] v(p_1),
\end{eqnarray}
\begin{eqnarray}
\delta M^{v(\hat{u})}=\delta M^{v(\hat{t})} (p_3 \leftrightarrow
p_4, \hat{t} \leftrightarrow \hat{u}),
\end{eqnarray}
\begin{eqnarray}
\delta M^{b(\hat{t})}&=&-i\frac{e^2}{16\pi^2} \epsilon_{\mu} (p_4)
\epsilon_{\nu}(p_3) \bar{u}(p_2) [f_1^{b(\hat{t})} \gamma^{\mu}
\gamma^{\nu} + f_2^{b(\hat{t})} \nonumber \\
&&  \times \gamma^{\nu}\gamma^{\mu}  + f_3^{b(\hat{t})} \gamma^{\mu}
p_1^{\nu}+ f_4^{b(\hat{t})} p_1^{\mu}
\gamma^{\nu}  + f_5^{b(\hat{t})} \gamma^{\mu} p_2^{\nu} \nonumber\\
&&+ f_6^{b(\hat{t})} p_2^{\mu} \gamma^{\nu} +f_7^{b(\hat{t})}
p_1^{\mu} p_1^{\nu} +f_8^{b(\hat{t})} p_1^{\mu} p_2^{\nu}+
f_9^{b(\hat{t})} \nonumber
\\ &&  \times p_2^{\mu} p_1^{\nu} +f_{10}^{b(\hat{t})}
p_2^{\mu} p_2^{\nu}
+f_{11}^{b(\hat{t})} \not p_4 \gamma^{\mu} \gamma^{\nu} +f_{12}^{b(\hat{t})} \nonumber\\
&&  \times \not p_4 \gamma^{\nu} \gamma^{\mu} +f_{13}^{b(\hat{t})}
\not p_4 \gamma^{\mu} p_1^{\nu} +f_{14}^{b(\hat{t})} \not p_4
p_1^{\mu} \gamma^{\nu} \nonumber \\ && +f_{15}^{b(\hat{t})} \not p_4
\gamma^{\mu} p_2^{\nu}
+f_{16}^{b(\hat{t})} \not p_4 p_2^{\mu} \gamma^{\nu} +f_{17}^{b(\hat{t})} \not p_4 \nonumber\\
&& \times p_1^{\mu} p_1^{\nu} +f_{18}^{b(\hat{t})} \not p_4
p_1^{\mu} p_2^{\nu} +f_{19}^{b(\hat{t})} \not p_4 p_2^{\mu}
p_1^{\nu} \nonumber \\ && +f_{20}^{b(\hat{t})} \not p_4 p_2^{\mu}
p_2^{\nu} ] v(p_1),
\end{eqnarray}
\begin{eqnarray}
\delta M^{b(\hat{u})}=\delta M^{b(\hat{t})} (p_3 \leftrightarrow
p_4, \hat{t} \leftrightarrow \hat{u}),
\end{eqnarray}
and
\begin{eqnarray}
\delta M^{\rm tr}&=&i\frac{1}{16\pi^2} \epsilon_{\mu} (p_4)
\epsilon_{\nu}(p_3) \bar{u}(p_2) [f_1^{\rm tr} \gamma^{\mu}
\gamma^{\nu}  \nonumber \\
&& + f_2^{\rm tr} \gamma^{\nu}\gamma^{\mu}+ f_3^{\rm tr}
\gamma^{\mu} p_1^{\nu}+ f_4^{\rm tr} p_1^{\mu}
\gamma^{\nu} \nonumber\\
&& + f_5^{\rm tr} \gamma^{\mu} p_2^{\nu} + f_6^{\rm tr} p_2^{\mu}
\gamma^{\nu} ] v(p_1).
\end{eqnarray}
Here $\hat{t}=(p_4-p_2)^2$, $\hat{u}=(p_4-p_1)^2$, $p_3$ and $p_4$
denote the momenta of the two incoming photons, and $p_2$ and
$p_1$ are the momenta of the outgoing bottom quark and its
antiparticle.

The form factors $f_i^{s(\hat{t})}$, $f_i^{v(\hat{t})}$,
$f_i^{b(\hat{t})}$ and $f_i^{\rm tr}$ are expressed in terms of
two-, three-, and four- point scalar integrals \cite{Clements1983},
and their analytical expressions are tedious, so we do not present
them. We can find that all the ultraviolet divergences cancel in the
form factors.

The cross section of the subprocess $\gamma\gamma \rightarrow
b\bar{b}$ for the unpolarized photons is given by
\begin{equation}
\hat{\sigma}(\hat{s})=\frac{N_C}{16\pi\hat{s}^2}
\int_{\hat{t}^-}^{\hat{t}^+} {\rm d} \hat{t} \overline{\sum_{\rm
spins}}|M_{\rm ren}(\hat{s},\hat{t})|^2,
\end{equation}
where
\begin{equation}
\hat{t}^{\pm}=(m_b^2-\frac{1}{2}\hat{s}) \pm \frac{1}{2}\hat{s}
\sqrt{1-4 m_b^2/\hat{s}}.
\end{equation}
The bar over the sum recalls averaging over initial spins and
\begin{equation}
\overline{\sum_{\rm spins}}|M_{\rm ren}(\hat{s},\hat{t})|^2=
\overline{\sum_{\rm spins}}|M_{0}|^2+2{\rm Re} \overline{\sum_{\rm
spins}}M_0^{\dagger}\delta M.
\end{equation}

The total cross section $\sigma(s)$ for the bottom pair production
can be obtained by folding the elementary cross section
$\sigma(\hat{s})$ for the subprocess $\gamma\gamma \rightarrow
b\bar{b}$ with the photon luminosity at the $e^+e^-$ colliders given
in Refs. \cite{Eboli1993} and \cite{Han1996}, i.e.,
\begin{equation}
\sigma(s)=\int_{2m_b/\sqrt{s}}^{x_{\rm max}} {\rm d}z \frac{{\rm
d}L_{\gamma\gamma}}{{\rm d}z}\hat{\sigma}(\hat{s}) \ \ (\gamma\gamma
\rightarrow b\bar{b} \ {\rm at} \ \hat{s}=z^2 s),
\end{equation}
where $\sqrt{s}$ and $\sqrt{\hat{s}}$ are the $e^+e^-$ and
$\gamma\gamma$ center-of-mass energies respectively, and ${\rm
d}L_{\gamma\gamma}/{\rm d}z$ is the photon luminosity, which can be
expressed as
\begin{equation}
 \frac{{\rm
d}L_{\gamma\gamma}}{{\rm d}z}=2z\int_{z^2/x_{\rm max}}^{x_{\rm max}}
\frac{{\rm d}x}{x} F_{\gamma/e}(x)F_{\gamma/e}(z^2/x).
\end{equation}
For unpolarized initial electron and laser beams, the energy
spectrum of the backscattered photon is given by
\cite{Eboli1993,Cheung1993}
\begin{equation}
 F_{\gamma/e}(x)=\frac{1}{D(\xi)}[1-x+\frac{1}{1-x}
 -\frac{4x}{\xi(1-x)}+\frac{4x^2}{\xi^2
 (1-x^2)}],
\end{equation}
with
\begin{equation}
 D(\xi)=(1-\frac{4}{\xi}-\frac{8}{\xi^2}){\rm ln}(1+\xi) +\frac{1}{2}
 +\frac{8}{\xi}-\frac{1}{2(1+\xi)^2},
\end{equation}
where $\xi=4E_e E_0/m_e^2$ in which $m_e$ and $E_e$ denote
respectively the incident electron mass and energy, $E_0$ denotes
the initial laser photon energy, and $x=E/E_e$ is the fraction which
represents the ratio between the scattered photon and initial
electron energy for the backscattered photons moving along the
initial electron direction. $F_{\gamma/e}(x)$ vanishes for $x>x_{\rm
max}=E_{\rm max}/E_e=\xi/(1+\xi)$. In order to avoid the creation of
$e^+e^-$ pairs by the interaction of the incident and backscattered
photons, we require $E_0 x_{\rm max} \leq m_e^2/E_e$ which implies
$\xi \leq 2+2\sqrt{2} \approx 4.8$ \cite{Cheung1993}. For the choice
$\xi=4.8$, it can obtain
\begin{equation}
x_{\rm max} \approx 0.83, \ \ D(\xi) \approx 1.8.
\end{equation}

\section{\label{sec:level4} Numerical results and conclusions}

There are several free parameters in the LHT model which are
involved in the amplitude of $\gamma\gamma \rightarrow b\bar{b}$.
They are the breaking scale $f$, the masses of the mirror quarks
$m_{Hi}$ $(i = 1, 2, 3)$ (here we have ignored the mass difference
between up-type mirror quarks and down-type mirror quarks at the
order up to $O(v/f)$), the mixing parameter $x_L$ between the SM top
quark and its heavy partner $T_+$ quark, and the other 6 parameters
($\theta^d_{12}, \theta^d_{13}, \theta^d_{23}, \delta^d_{12},
\delta^d_{13}, \delta^d_{23}$), which are related to the mixing
matrix $V_{Hd}$.

For the parameters $f$ and $x_L$, some constraints come from the
electroweak precision measurements and the WMAP experiment for
dark matter relics \cite{Matsumoto2008}, which shows that the
region $f< 570\ {\rm GeV}$ is kinematically forbidden. However,
these constraints also depend on the other parameters. Hence, we
slightly relax
 the constraints on the parameters $f$ and $x_L$, and let
them vary in the range
\begin{equation}
500\ {\rm GeV} \leq  f \leq 1500\ {\rm GeV}, \ \ 0.1 \leq x_L \leq
0.8,
\end{equation}
in our numerical calculations.

In Refs. \cite{Hubisz2006,Blanke2006}, the constraints on the mass
spectrum of the mirror fermions have been investigated from the
analysis of neutral meson mixing in the $K$, $B$ and $D$ systems. It
has been found that a TeV scale GIM suppression is necessary for a
generic choice of $V_{Hd}$. However, there are regions of parameter
space which are only very loose constraints on the mass spectrum of
the mirror fermions. For the matrix $V_{Hd}$, we follow Ref.
\cite{Hou2007} to consider two scenarios for these parameters to
simplify our calculations:

(I) $V_{Hd}=1$. This scenario is  connected only with the
third-generation mirror quarks due to its involvement in bottom
quark and its antiparticle in the final states. Moreover, the
constraints on the mass spectrum of the mirror fermions can be
relaxed \cite{Hubisz2006}. Therefore, we take
\begin{equation}
500\ {\rm GeV}\leq m_{H3} \leq 3000\ {\rm GeV},
\end{equation}
to see its effect.

(II) $s^d_{23}=1/\sqrt{2}$, $s^d_{12}=s^d_{13}=0$,
$\delta^d_{12}=\delta^d_{23}=\delta^d_{13}=0$. In this scenario, the
$D$-meson system can give strong constraints on the relevant
parameters \cite{Hubisz2006}. Considering these constraints, we fix
$m_{H_1}=m_{H_2}=500\ {\rm GeV}$, and take the same assumption as in
Scenario-I for the third-generation mirror quarks.

\begin{figure}
\includegraphics{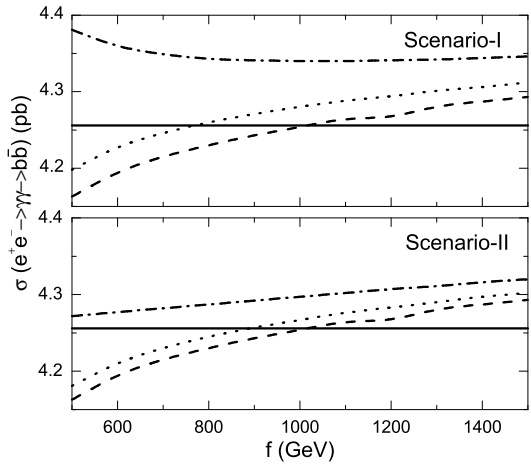}
\caption{\label{fig:eps2} The total cross section $\sigma (e^+e^-
\rightarrow \gamma\gamma \rightarrow b\bar{b})$ as a function of $f$
for $\sqrt{s} = 500$ GeV and $x_L = 0.3$. The dashed lines, dotted
lines and dot-dashed lines denote, respectively, the cases of
$m_{H3}=500, 1000, 1500\ {\rm GeV}$, and the solid lines stand for
the results of the tree level.}
\end{figure}

\begin{figure}
\includegraphics{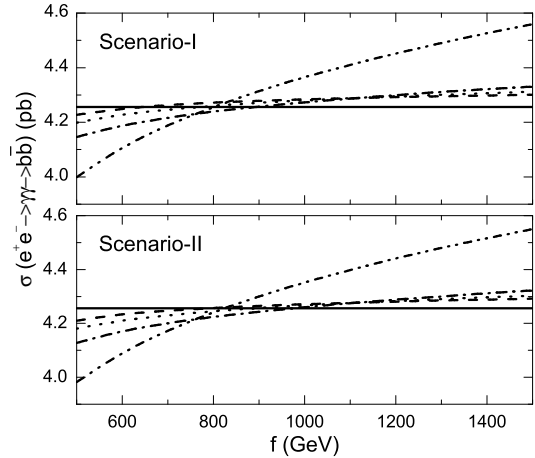}
\caption{\label{fig:eps3} The total cross section $\sigma (e^+e^-
\rightarrow \gamma\gamma \rightarrow b\bar{b})$ versus $f$ with
$\sqrt{s} = 500$ GeV and $m_{H3}=1000\ {\rm GeV}$. The dashed lines,
dotted lines, dot-dashed lines and  dot-dot-dashed lines indicate
the cases of $x_L =0.1, 0.3$, $0.5$ and $0.8$, respectively, and the
solid lines represent the results of the tree level.}
\end{figure}

\begin{figure}
\includegraphics{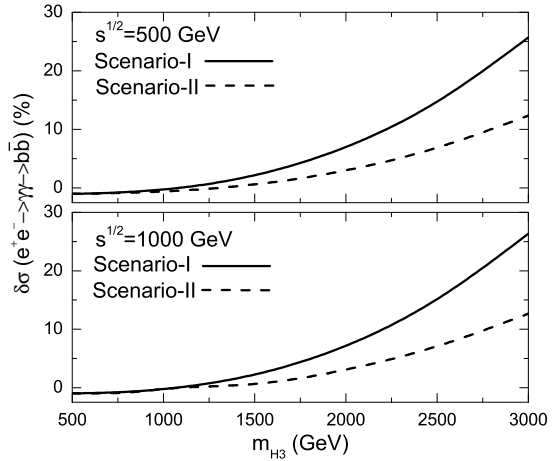}
\caption{\label{fig:eps4} The relative correction of the LHT model
to the cross section, $\delta \sigma (e^+e^- \rightarrow
\gamma\gamma \rightarrow b\bar{b})$, as a function of $m_{H3}$
with $f=700$ GeV and $x_L = 0.3$.}
\end{figure}

In our numerical evaluation, we take a set of independent input
parameters which are known from current experiment. The input
parameters are $m_t=171.2\ {\rm GeV}$, $m_b=4.2\ {\rm GeV}$, $M_W=
80.398\ {\rm GeV}$, $M_Z=91.1876\ {\rm GeV}$, $\alpha=1/137.036$ and
$G_F=1.16637 \times 10^{-5}\ {\rm GeV^{-2}}$ \cite{Amsler2008}. For
the c.m. energies of the International Linear Collider (ILC), we
choose $\sqrt{s}=500, 1000\ {\rm GeV}$ according to the ILC
Reference Design Report \cite{Brau2007}. The final numerical results
are summarized in Figs.~\ref{fig:eps2}-\ref{fig:eps4}.

Figure \ref{fig:eps2} shows the total cross section $\sigma (e^+e^-
\rightarrow \gamma\gamma \rightarrow b\bar{b})$ versus $f$ with
$\sqrt{s} = 500$ GeV and $x_L = 0.3$, in which the dashed lines,
dotted lines and dot-dashed lines denote the cases of $m_{H3}=500,
1000, 1500\ {\rm GeV}$, respectively, and the solid lines stand for
the results of the tree level. From this figure, we can obtain the
following results: (i) The cross section is strongly dependent on
the parameter $f$, and is about a few percent to dozens of percent
larger than that of the tree level. It is natural since the
couplings between the new top quark $T_+$ and the SM quarks are
proportional to the mass of $T_+$ quark; i.e., are proportional to
the breaking scale $f$. Furthermore, our analytical calculations
also show that the contributions from the heavy T-odd gauge bosons
and Goldstone bosons increase slightly with $f$ for $m_{H3} \leq
1000\ {\rm GeV}$, and decrease slowly when $m_{H3} > 1000\ {\rm
GeV}$; (ii) Since the couplings between the mirror quarks and the SM
quarks are proportional to the mirror quark masses, the cross
section increases distinctly with $f$ for the cases of $m_{H3}$ from
$500$, $1000$ to $1500$ GeV, while the relative section cross is
negative when all of $f$ and $m_{H3}$ take small values; and (iii)
Comparing these two scenarios, we can see that the cross section for
Scenario-II does not have a large deviation from that for Scenario-I
when $m_{H3}$ takes a small value, but the former is only about a
half of the latter for a large value of $m_{H3}$.

The cross section $\sigma (e^+e^- \rightarrow \gamma\gamma
\rightarrow b\bar{b})$ versus the parameter $f$ for various values
$x_L$ when $\sqrt{s} = 500$ GeV and $m_{H3}=1000\ {\rm GeV}$ is
given in Fig.~\ref{fig:eps3}.  We can see that: (i) the correction
of the LHT model to the cross section changes from negative to
positive with $f$ becoming large; (ii) the increment of the cross
section with $f$ is slow for the cases of $x_L =0.1, 0.3$, $0.5$ and
is quick for $x_L=0.8$; and (iii) the behavior of the cross section
for Scenario-II is almost the same as that for Scenario-I.

For the case of $\sqrt{s}=1000$ GeV, our calculations show that
the effect of the LHT model in this case is slightly larger than
that in the case of $\sqrt{s}=500$ GeV.

In order to look at the relative correction of the LHT model to the
cross section, we take $f=700$ GeV and $x_L = 0.3$ as an example and
plot $\delta\sigma(e^+e^- \rightarrow \gamma\gamma \rightarrow
b\bar{b})$ as a function of $m_{H3}$ in Fig.~\ref{fig:eps4}. From
this figure, we can find that (i) the contribution of the LHT model
to the process is very obvious unless $m_{H3}$ is small, (ii) for
the case of $\sqrt{s}=500$ GeV, the relative correction,
$\delta\sigma(e^+e^- \rightarrow \gamma\gamma \rightarrow
b\bar{b})$, is sensitive to $m_{H3}$, and increases with $m_{H3}$
from $-0.97\%$ to $25.71\%$ for Scenario-I and from $-0.97\%$ to
$12.37\%$ for Scenario-II, and (iii) for $\sqrt{s}=1000$ GeV, the
relative correction changes from $-0.99\% \sim 26.32\%$ for
Scenario-I, and $-0.99\% \sim 12.66\%$ for Scenario-II.

We know that the ILC is the important next generation linear
collider. According to the ILC Reference Design Report
\cite{Brau2007}, the ILC is determined to run with $\sqrt{s}=500$
GeV (upgradeable to $1000$ GeV) and the total luminosity required is
$L=500\ {\rm fb}^{-1}$ with the first four year operation and
$L=1000\ {\rm fb}^{-1}$ during the first phase of operation with
$\sqrt{s}=500$ GeV. It means that, millions of the bottom pairs per
year can be produced, and the relative correction of the LHT model
to the cross section can reach the level from a few percent to
dozens of percent when $m_{H3}$ takes a larger value. However, the
relative correction induced by the charged Higgs and charged
Goldstone bosons in the 2HDM and MSSM is less than $0.1\%$
\cite{Han1996}, and in the TC2 model, the relative correction from
the pseudo-Goldstone bosons and the new gauge bosons is negative and
no more than $10\%$ \cite{Huang2008}. Furthermore, our calculations
show that the contribution of Higgs boson in the SM is only the
order of $10^{-6}$ which  is negligibly small. Therefore via the
process $e^+e^-\rightarrow \gamma\gamma \rightarrow b\bar{b}$, the
LHT model is experimentally distinguishable from the SM, 2HDM, MSSM
and TC2 models, which affords the possibility to test the LHT model
at the ILC unless $u_H^3$ and $d_H^3$ are very light. It is hoped
that ILC will be able to give strong constraints on the relevant
parameters of LHT model since the correction of the LHT model to the
cross section of $e^+e^- \rightarrow \gamma\gamma \rightarrow
b\bar{b}$ is sensitive to some parameters.

In conclusion, we have studied  the contribution of the LHT model to
the process $e^+e^- \rightarrow \gamma\gamma \rightarrow b\bar{b}$.
We find that, for the favorable parameters, the total cross section
$\sigma(e^+e^- \rightarrow \gamma\gamma \rightarrow b\bar{b})$ is
sensitive to the breaking scale $f$, the mixing parameter $x_L$, the
masses of the mirror quarks $m_{Hi}$, and the relative correction of
the LHT model is a few percent to dozens of percent unless $m_{H3}$
is very small. The total cross section is significantly larger than
the corresponding results in the standard model, the general two
Higgs doublet model, the minimal supersymmetric standard model, and
the topcolor assisted technicolor model. Therefore the difference is
obvious for the International Linear Colliders and it is really
interesting in testing the standard model and searching for the
signs of the littlest Higgs model with $T$ parity.

\begin{acknowledgments}

This project is supported in part by the Natural Science Foundation
of Henan Province under No. 0611050300; the Ph.D Programs Foundation
of Ministry of Education of China under No. 20060476002; the
National Natural Science Foundation of China under Grant Nos.
10575029, 10775039 and 10847120; and the Project of Knowledge
Innovation Program (PKIP) of Chinese Academy of Sciences under Grant
No. KJCX2.YW.W10.

\end{acknowledgments}

\end{document}